\documentclass[onecolumn,preprint,preprint,showpacs,12pt]{elsarticle}
\usepackage{bm}
\usepackage{amsmath,amsfonts}
\usepackage{siunitx}
\usepackage{graphicx}
\usepackage{amssymb}
\usepackage{lineno}
\usepackage[english]{babel}
\usepackage{xcolor}
\newcommand{\upcite}[1]{\cite{#1}}

\biboptions{numbers,sort&compress}

\journal{Matter and Radiation at Extremes}

\begin{document}

\begin{frontmatter}

\title{Radiation reaction induced spiral attractors in ultra-intense colliding laser beams}

\author[label1]{Zheng Gong}
\author[label1]{Ronghao Hu}
\author[label1]{Yinren Shou}
\author[label1]{Bin Qiao}
\author[label1]{Chiaer Chen}
\author[label1]{Furong Xu}
\author[label1]{Xiantu He}
\author[label1,label2]{Xueqing Yan\footnotetext{$\dagger$ x.yan@pku.edu.cn}}
\address[label1]{State Key Laboratory of Nuclear Physics and Technology, and Key Laboratory of HEDP of the Ministry of Education, CAPT, Peking University, Beijing, China, 100871}
\address[label2]{Collaborative Innovation Center of Extreme Optics, Shanxi University, Taiyuan, Shanxi, China, 030006}
\date{\today}%

\begin{abstract}
The radiation reaction effects on electron dynamics in counter-propagating circularly polarized laser beams are investigated through the linearization theorem and the results are in great agreement with numeric solutions. For the first time, the properties of fixed points in electron phase-space were analyzed with linear stability theory, showing that center nodes will become attractors if the classical radiation reaction is considered. Electron dynamics are significantly affected by the properties of the fixed points and the electron phase-space densities are found to be increasing exponentially near the attractors. The density growth rates are derived theoretically and further verified by particle-in-cell simulations, which can be detected in experiments to explore the effects of radiation reaction qualitatively. The attractor can also facilitate to realize a series of nanometer-scaled flying electron slices via adjusting the colliding laser frequencies.
\end{abstract}
\begin{keyword}
 Radiation reaction effect \sep Phase space dynamics \sep PIC simulation
 \PACS 52.38.-r \sep 05.45.-a \sep 41.85.Ct
\end{keyword}
\end{frontmatter}

\section{Introduction}
\label{Section1}
The interactions of ultra-short and ultra-intense laser pulses with various plasmas can generate brilliant sources of energetic electrons, ions, x/$\gamma$-rays, and positrons with proper laser plasma parameters\upcite{RevModPhys.78.309,RevModPhys.81.1229,RevModPhys.85.751,RevModPhys.85.1,di2012extremely}. Electrons are the most fundamental particles in laser plasma interaction as electrons can be easy accelerated to relativistic velocities with laser intensities higher than \SI{e18}{W/cm^2}\upcite{gibbon2004short,meyer2001relativistic,Toncian201682}. The electron dynamics in laser fields has been investigated thoroughly under the framework of classical electrodynamics. However, with the advent of more powerful laser facilities, laser intensities are about to achieve \SI{e23}{W/cm^2}\upcite{mourou2011more} and electron dynamics in such intense laser fields are substantially different since here the magnitude of radiation reaction (RR) force and Lorentz force are comparable\upcite{bulanov2016relativistically}. Quantum electrodynamics (QED) based numeric method\upcite{duclous2011monte,Ridgers2014273,bell2008possibility,PRL2010Di_Piazza,PRL2013Neitz} provides a explicitly self-consistent description of electron discrete emission and the corresponding radiation recoil. Semi-classical description\upcite{Sokolov2009Dynamics,PhysRevLett.108.195001} of radiation reaction force provides a reliable theoretical method to estimate the continuous radiation effects, avoiding the well-known self-acceleration solutions of classical models\upcite{jackson1999classical}. Novel phenomena beyond the framework of classical electrodynamics are predicted by the semi-classical method and QED model such as the radiation trapping\upcite{PhysRevLett.112.145003,PhysRevLett.113.014801,esirkepov2014attractors},  phase space contraction\upcite{lehmann2012phase,tamburini2011radiation}, QED induced stochastic effect\upcite{PRL2013Neitz,vranic2015quantum} and $e^+e^-$ pair production\upcite{bulanov2010multiple}.

Experimental detection of radiation reaction effect could be difficult as it is almost unaccessible to measure the microscopic quantities of a single electron motion. However, with proper experimental setup, the tiny differences in electron dynamics can lead to the changes of macroscopic quantities that can be measured with available techniques\upcite{PhysRevLett.112.015001,PhysRevE.91.023207}.
Detecting the angular distribution changing of the electron or its emitted photon in counter-propagating laser fields provides a optimal method to qualitatively explore the signatures of radiation reaction\upcite{PhysRevX.2.041004,wang2015signatures}. Recently, $\gamma$-ray generation and pair production in counter-propagating laser fields have been investigated widely\upcite{nerush2011laser,Chang2015Generation,M2016Electron}, whereas the electron spatiotemporal evolution nearby the attractors in intense colliding laser beams still lacks a quantitative prediction. On the other hand, for laser radiation with 1$\mu$m wavelength the radiation friction force changes the scenario of the electromagnetic wave interaction with matter at the intensity of $I_R\approx$10$^{23}$W/cm$^2$. For the laser intensity close to $I_R$, the electron interaction with the electromagnetic field is principally determined by a counterplay between the radiation friction and quantum effects\upcite{di2012extremely,Bulanov2015}. When laser intensity is higher than $I_Q$=$5.75\times10^{23}(1\mu m/\lambda)$\,W/cm$^2$, the QED effects weaken the electromagnetic emission\upcite{kirk2009pair} and the process of photon emission becomes stochastic\upcite{duclous2011monte}. Provided the QED induced radiation weaken and stochastic effect are indifferent under the laser intensity lower than $I_Q$, the classical radiation reaction approach still gives us valid results. In this paper, we utilizing the classical radiation reaction model to investigate the electron dynamics in counter-propagating laser fields. The presented electron spatiotemporal evolution were dominantly affected by the fixed points in electron phase-space and the analytical solutions of electron dynamics were obtained with linear stability theory\upcite{hirsch2012differential}. It is found that the spiral attractors induced by radiation reaction can lead to exponential growth of in situ density and the analytic growth rates were given and compared with numeric solutions.

\section{Theoretical analysis}
\label{Section2}
For simplicity and without losing generality, the counter propagating laser pulses are described by infinite plane wave vector potential, $\mathbf{A}_1=a_0[sin(t-x)\hat{y}+cos(t-x)\hat{z}]$ and $\mathbf{A}_2=a_0[sin(t+x)\hat{y}+cos(t+x)\hat{z}]$; $\hat{y}\ (\hat{z})$ is the unit vector in $y\ (z)$ direction; $a_{0}$ is the normalized laser amplitude ($a_{0}=eE_{0}/m_{e}cw_0$), where $e$ and $m_{e}$ are the electron charge and mass, $E_{0}$ and $w_{0}$ are the electric amplitude and frequency, c is the speed of light, respectively; $x$ and $t$ are normalized to $c/w_0$ and $w_0$. The electromagnetic standing wave (SW) field can be deduced as $\mathbf{E}=-\partial\mathbf{A}/\partial t=-2a_0cos(x)cos(t)\hat{y}+2a_0cos(x)sin(t)\hat{z}$ and $\mathbf{B}=\nabla\times\mathbf{A}=2a_0sin(x)cos(t)\hat{y}-2a_0sin(x)sin(t)\hat{z}$ from the whole region vector potential $\mathbf{A}=\mathbf{A}_1+\mathbf{A}_2$.

\begin{figure}[!]
\centering
\includegraphics[width=\textwidth]{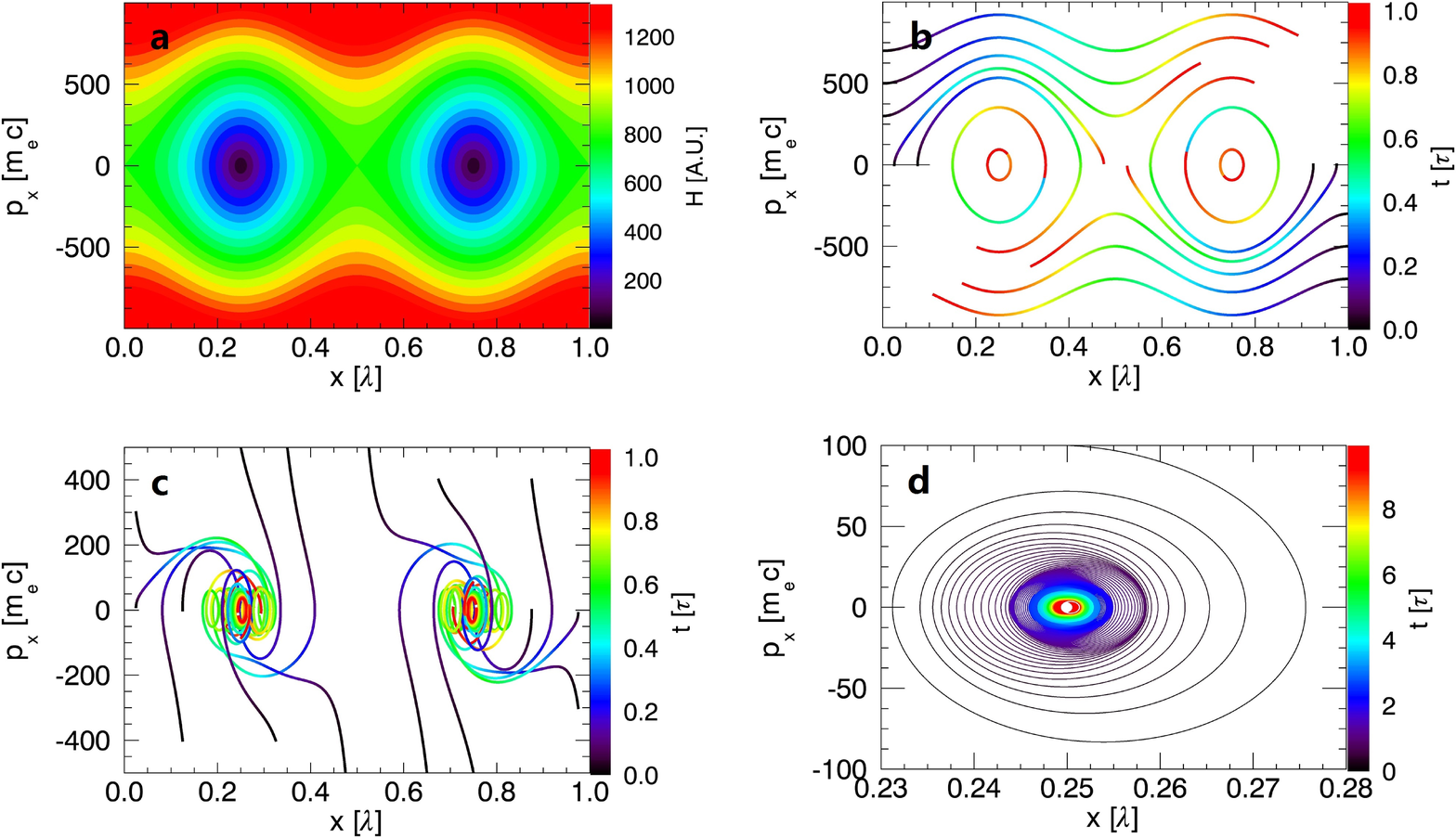}
\caption{Equation integration result at laser field amplitude $a_0=300$. (a) The conservative Hamiltonian for electron satisfying initial condition $p_y|_{t=0}=A_y|_{t=0}$ ($p_z|_{t=0}=A_z|_{t=0}$), with RR effect ignored. For denotion simplicity the unite in x-direction is laser wavelength $\lambda$ rather than inverse of wave number $\frac{\lambda}{2\pi}$. (b) Trajectories of electron in phase-space $(x,p_x)$ without RR, which illustrates the node $(x=0.25\lambda\ or\ 0.75\lambda)$ acts as a center. Color denotes the temporal variation in laser period $\tau$. (c) Same as (b), but RR effect included, where node behaves like a attractor but the spiral phenomenon doesn't exhibit. (d) The electron approaches to node spirally when it is close to the node}
\label{fig1}
\end{figure}

First of all, considering the condition without RR recoil, electron dynamics in phase-space $x-p_x$ is determined by relativistic Lorentz force $d\mathbf{p}/dt=\partial\mathbf{A}/\partial t-\mathbf{v}\times(\nabla\times\mathbf{A})$. As SW field is independent of $y$ and $z$, there are two invariant equations:
\begin{equation}
\begin{aligned}\label{eq1}
 \frac{dp_y}{dt}=\frac{\partial A_y}{\partial t}+v_x\frac{\partial A_y}{\partial x}=\frac{dA_y}{dt}
 \end{aligned}
\end{equation}
\begin{equation}
\begin{aligned}\label{eq2}
 \frac{dp_z}{dt}=\frac{\partial A_z}{\partial t}+v_x\frac{\partial A_z}{\partial x}=\frac{dA_z}{dt}
 \end{aligned}
\end{equation}
Assuming at initial time $p_y|_{t=0}=A_y|_{t=0}$ ($p_z|_{t=0}=A_z|_{t=0}$), above conservative relation tell us $p_z\equiv A_z=2a_0cos(x)cos(t)$ ($p_y\equiv A_y=2a_0cos(x)sin(t)$). The relativistic factor $\gamma=\sqrt{1+{p_x}^2+4{a_0}^2cos^2(x)}$ and the nonlinear differential equation in $x-p_x$ space is derived:
\begin{equation}
\begin{aligned}\label{eq3}
 \frac{dx}{dt}=\frac{p_x}{\gamma}=\frac{p_x}{\sqrt{1+{p_x}^2+4{a_0}^2cos^2(x)}}
 \end{aligned}
\end{equation}
\begin{equation}
\begin{aligned}\label{eq4}
 \frac{dp_x}{dt}=-v_yB_z+v_zB_y=\frac{4{a_0}^2cos(x)sin(x)}{\sqrt{1+{p_x}^2+4{a_0}^2cos^2(x)}}
 \end{aligned}
\end{equation}
From the time-independent nonlinear relationship Eq.(3)(4) (i.e., autonomous nonlinear system in mathematics), we can exactly find the relativistic factor is a conservative Hamiltonian
\begin{equation}
\begin{aligned}\label{eq5}
 \textit{H}=\gamma=\sqrt{1+{p_x}^2+4{a_0}^2cos^2(x)}
 \end{aligned}
\end{equation}
since $d\textit{H}/dt=\partial\textit{H}/\partial t\equiv0$ is validated from the corresponding canonical equation $dx/dt=\partial \gamma/\partial p_x=f(x,p_x)$ and $dp_x/dt=-\partial \gamma/\partial x=g(x,p_x)$ which are completely equivalent with Eq.(3)(4). The Hamiltonian $\textit{H}$ is symmetrical and periodic, and note that there are some special solution of these differential equation when initial value $(x^*,p_x^*)$ satisfies $f(x^*,p_x^*)=0$ and $g(x^*,p_x^*)=0$. These are the constant solution $(x,p_x)\equiv(x^*,p_x^*)$. A constant solution such as these is called an equilibrium solution or equilibrium point for the equation\cite{hirsch2012differential}. Subsequently there are four equilibrium points $(x^*,p_x^*)$ at electric node $(\pi/2,0),\ (3\pi/2,0)$ and antinode $(0,0),\ (\pi,0)$ in a SW period, as shown in Fig.\ref{fig1}(a). The property of an equilibrium point in the nonlinear system can be classified via its linear approximation nearby the equilibrium point. The Jacobian matrix is a linearization method via calculating first partial derivatives which facilitates us to investigate the property of the equilibrium point\cite{jordan2007nonlinear,hirsch2012differential}. To determine whether the equilibrium point is stable or not, making disturbance expansion nearby $(x-x^*,p_x-p_x^*)$ and dropping quadratic terms to linearize Eq.(3)(4), the characteristic Jacobian matrix $\mathbf{Ja}$ at the equilibrium point $(x^*,p_x^*)$ is obtained:
\begin{equation}
\begin{aligned}\label{eq6}
 \mathbf{Ja}=\begin{pmatrix} \frac{\partial f(x,p_x)}{\partial x} & \frac{\partial f(x,p_x)}{\partial p_x} \\ \frac{\partial g(x,p_x)}{\partial x} & \frac{\partial g(x,p_x)}{\partial p_x} \end{pmatrix}_{x^*,p_x^*}=\begin{pmatrix} 0 & \frac{1}{\gamma} \\ \frac{4a_0^2cos(2x)}{\gamma} & 0 \end{pmatrix}_{x^*,p_x^*}
 \end{aligned}
\end{equation}
For electric node $x^*=\pi/2\ or\ 3\pi/2$, the trace and determinant of Jacobian matrix are $tr(Ja)=0$ and $det(Ja)=4a_0^2>0$, which manifests $(\pi/2,0),\ (3\pi/2,0)$ is a center without any source or sink effect when excluding RR\cite{hirsch2012differential}. The counterparts at $x^*=0\ or\ \pi$, $tr(Ja)=0$ and $det(Ja)=-\frac{4a_0^2}{1+4a_0^2}<0$, indicates antinode $(0,0),\ (\pi,0)$ is a unstable saddle point where tiny disturbance can trigger the drastic turbulence growth. In Fig.\ref{fig1}(b), the orbits of electron without RR, initially satisfying condition $p_\perp|_{t=0}=A_\perp|_{t=0}$, are attained by integrating Eq.\ref{eq3} and Eq.\ref{eq4}. This analysis method with respect to property of equilibrium point can be carried forward to RR circumstance directly.

On account of reasonability and causality of the RR effect, in the paper we use the Landau Lifshitz(LL) force\cite{landau1975lifshitz} for RR form to explore its influence on the equilibrium point. The spatial components of LL force are:
\begin{equation}
\begin{split}\label{eq7}
   \mathbf{F_{LL}}= & -\frac{4\pi r_e}{3\lambda}\{\gamma[(\frac{\partial}{\partial t}+\mathbf{v}\cdot\nabla)\mathbf{E}+\mathbf{v}\times(\frac{\partial}{\partial t}+\mathbf{v}\cdot\nabla)\mathbf{B}]\} \\
     & +\frac{4\pi r_e}{3\lambda}\{[(\mathbf{E}+\mathbf{v}\times\mathbf{B})\times\mathbf{B}+(\mathbf{v}\cdot\mathbf{E})\mathbf{E}]-\gamma ^2[(\mathbf{E}+\mathbf{v}\times\mathbf{B})^2-(\mathbf{v}\cdot\mathbf{E})^2]\mathbf{v}\}
\end{split}
\end{equation}
where $\mathbf{p}$ is the electron momentum, $r_e\equiv e^2/mc^2\approx2.8\times10^{-15} m$ is the classical electron radius, $\lambda=2\pi c/w_0$ is the pulse wavelength and we use dimensionless quantities for convenience. Here momentum in units of $mc$, electric and  magnetic field in units of $mw_0c/e$ and $mw_0c^2/e$ respectively. The first term of Eq.(7), the one containing the 'total' time derivative of the electromagnetic field is neglected in calculation. Since RR usually are considered in ultra-relativistic condition, $\gamma\gg1$, the last term (proportional to $\gamma^2$) dominates over the whole LL force. Analogous to the case without RR, the corresponding Jacobian matrix with $\mathbf{F_{LL}}$ is estimated as:

\begin{equation}
\begin{aligned}\label{eq8}
 \mathbf{Ja}\approx\begin{pmatrix} 0 & \frac{1}{\gamma} \\ \frac{4a_0^2cos(2x)}{\gamma} & -\frac{4\pi r_e}{3\lambda}2a_0^2\gamma \end{pmatrix}_{x^*,p_x^*}
 \end{aligned}
\end{equation}

Using the classification principle in trace-determinant plane, antinode $(0,0)$ or $(\pi,0)$ still behaves the same unstable saddle point as no RR case. However, at the node $(\pi/2,0)$ or $(3\pi/2,0)$, $tr(Ja)=-\frac{4\pi r_e}{3\lambda}2a_0^2\sqrt{1+4a_0^2}<0$ and $det(Ja)=4a_0^2>0$, the node's behaviour converts from center to spiral sink when RR force is taken into account. As shown in Fig.\ref{fig1}(c), the electrons initially far from the node are attracted by it violently while they experiencing swift movement in phase-space. This can be seen as electrons can no longer stay in the original height when friction (RR) exerts on them to dissipate their energy and eventually they fall to valley bottom (spiral sink node). Since the RR force in the above model where electron initial far from the equilibrium node is comparable with Lorentz force at $a_0=300$ and $\gamma\approx500$ when considering estimation $\mathbf{F_{LL}}\sim-\frac{4\pi r_e}{3\lambda}\gamma^2a_0^2$, electrons move to the sink node swiftly rather than approach to the sink node spirally. The electron initially closed to node behaves spiral sink characteristic regularly in Fig.\ref{fig1}(d), which agrees well with analysis of Eq.\ref{eq8}.
Under one-order perturbation approximation, the dynamical differential equation at $(\tilde{x},\tilde{p_x})=(x-x^*,p_x-p_x^*)$ is expressed by linear algebra equation $\lambda X=AX$, here $X=(\tilde{x},\tilde{p_x})^\mathrm{T}$ and $A=\mathbf{Ja}$. The general solution $X=\alpha e^{\lambda_+t}X_1+\beta e^{\lambda_-t}X_2$ illustrates the electron trajectories in $x-p_x$ phase space, where $\lambda_{+(-)}$ is the eigenvalue of matrix A, $\alpha(\beta)$ is the undefined constant and $X_1(X_2)$ is the eigenvector. After some miscellaneous calculations, eigenvalues $\lambda_{\pm}=-2\frac{4\pi r_e}{3\lambda}\gamma a_0^2\pm2a_0i$ denotes the eigenvector's amplitude and the decreasing rate equals $\Re(\lambda_{\pm})=-2\frac{4\pi r_e}{3\lambda}\gamma a_0^2$. In the above series of derivation, we concentrate on integral variation presuming the system is autonomous and in consequence instantaneous variation of SW field $sin^2(t)$ and $cos^2(t)$ are replaced by 0.5 for period average effect. Previously Tambrini et al\cite{tamburini2011radiation} and G.Lehnman et al\cite{lehmann2012phase} qualitatively summarize the entropy differential $dS/dt=\int d^3xd^3pf\nabla_{p}\cdot\mathbf{F}_{LL}\leqslant0$ and the estimated characteristic attraction time $\sim[\frac{4\pi r_e}{3\lambda}\gamma a_0^2]^{-1}$\cite{lehmann2012phase} which are in good coincidence with our linearized derivation. Therefore considering the conservation of quantity and amplitude decay, the 2D phase volume contraction ratio is estimated as $e^{-2|\lambda_+\lambda_-|t}$.

\begin{figure}[!]
\centering
\includegraphics[width=\textwidth]{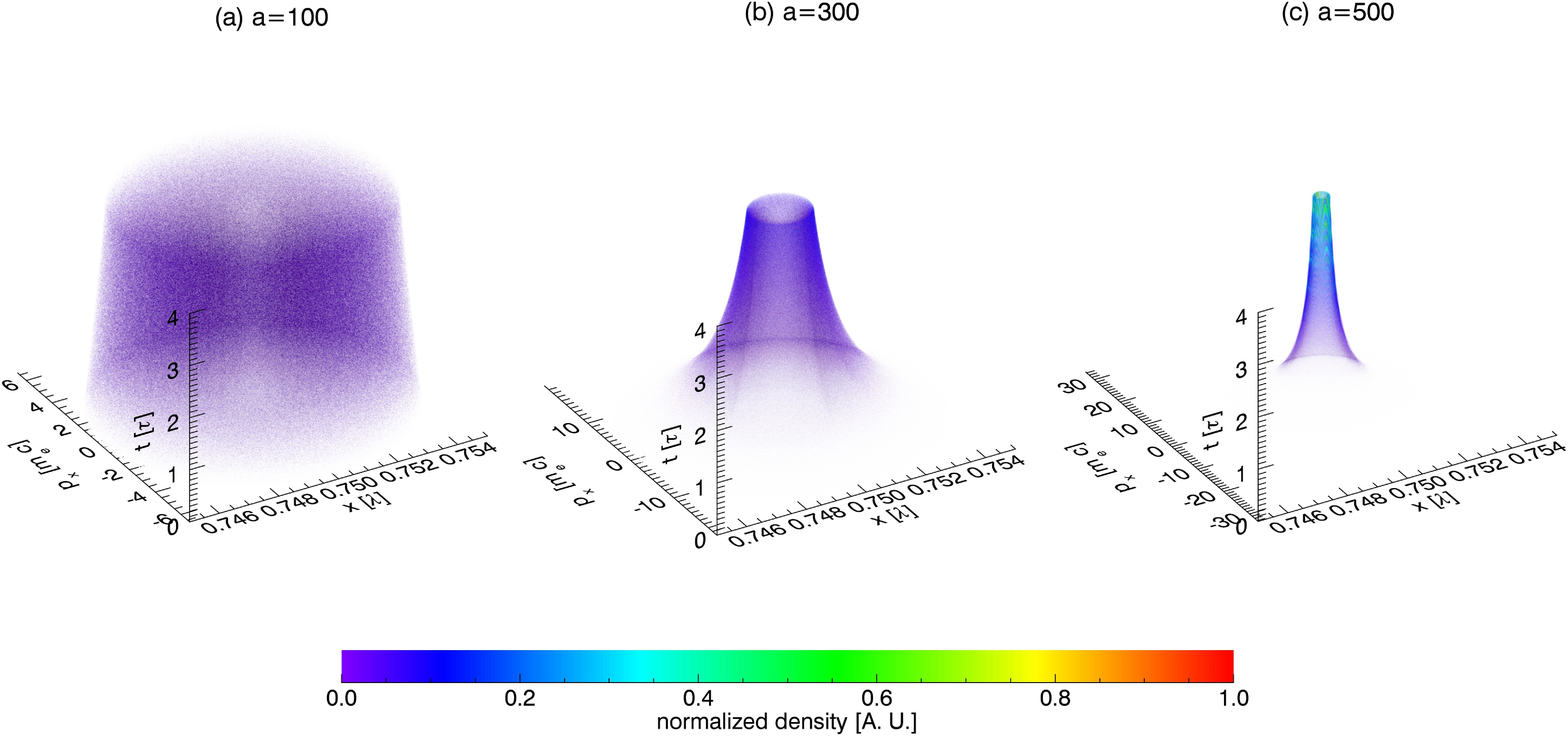}
\caption{(a) The electrons phase-space $(x,p_x)$ temporal evolution at normalized laser intensity $a_0=100$ during $0<t<4\tau$. (b)(c) the same as (a) with $a_0=300$ and $a_0=500$, respectively}
\label{fig2}
\end{figure}

\section{Simulation results}
\label{Section3}
To testify the numerical decay ratio, we take the advantage of one-dimension Particle in Cell (PIC) program EPOCH\cite{arber2015contemporary,ridgers2012dense} to simulate electron motion. The Lorentz force combined with LL force is realized in particle pusher model of EPOCH source code utilizing the method in Ref \cite{tamburini2010radiation}. The SW field $\mathbf{E}=-2a_0cos(\frac{2\pi}{\lambda}x)cos(\omega t)\hat{y}+2a_0cos(\frac{2\pi}{\lambda}x)sin(\omega t)\hat{z}$ and $\mathbf{B}=2a_0sin(\frac{2\pi}{\lambda}x)cos(\omega t)\hat{y}-2a_0sin(\frac{2\pi}{\lambda}x)sin(\omega t)\hat{z}$ is established by counter-propagation CP infinite plane wave. Meanwhile simulation region $x\in[0,\lambda]$ is divided into $10^5$ cell. As perturbation approximation method is valid nearby the node, we initialize $10^7$ electrons uniformly distributed in $(\frac{p_x}{a_0m_ec})^2+(\frac{4\pi(x-0.75\lambda)}{\lambda})^2\leqslant(0.02\pi)^2$ to explore the phase volume contraction ratio. Supposing electron bunch is so rare that space charge interaction is incomparable with SW field, collective force is neglected in our PIC simulation by setting the electron as the tracer particle. As shown in Fig.\ref{fig2}, RR induced phase space evolution during $0<t<4\tau$ is exhibited for different SW amplitude $a_0=100,300,500$, where $\tau=2\pi/\omega$ is laser period. The derived ratio $e^{-2|\lambda_+\lambda_-|t}=e^{-\frac{16\pi r_e}{3\lambda}\gamma a_0^2t}$ implies contraction is sensitive with amplitude $a_0$. The $a_0=500$ case illustrates electrons are immediately trapped to attractor (spiral node $(0,0.75\lambda)$) and they are accumulated to extreme dense state after several laser periods. However, under $a_0=100$ circumstance the contraction phenomenon is negligible and this bunch still distributes in a relative large elliptical region at $t=4\tau$. In-between the phase-space variation is observable but not so drastic for $a_0=300$ which is in accordance with precedent predicted threshold of radiation reaction dominant regime in Ref\cite{Bulanov2015,di2012extremely}.

\begin{figure}[!]
\centering
\includegraphics[width=\textwidth]{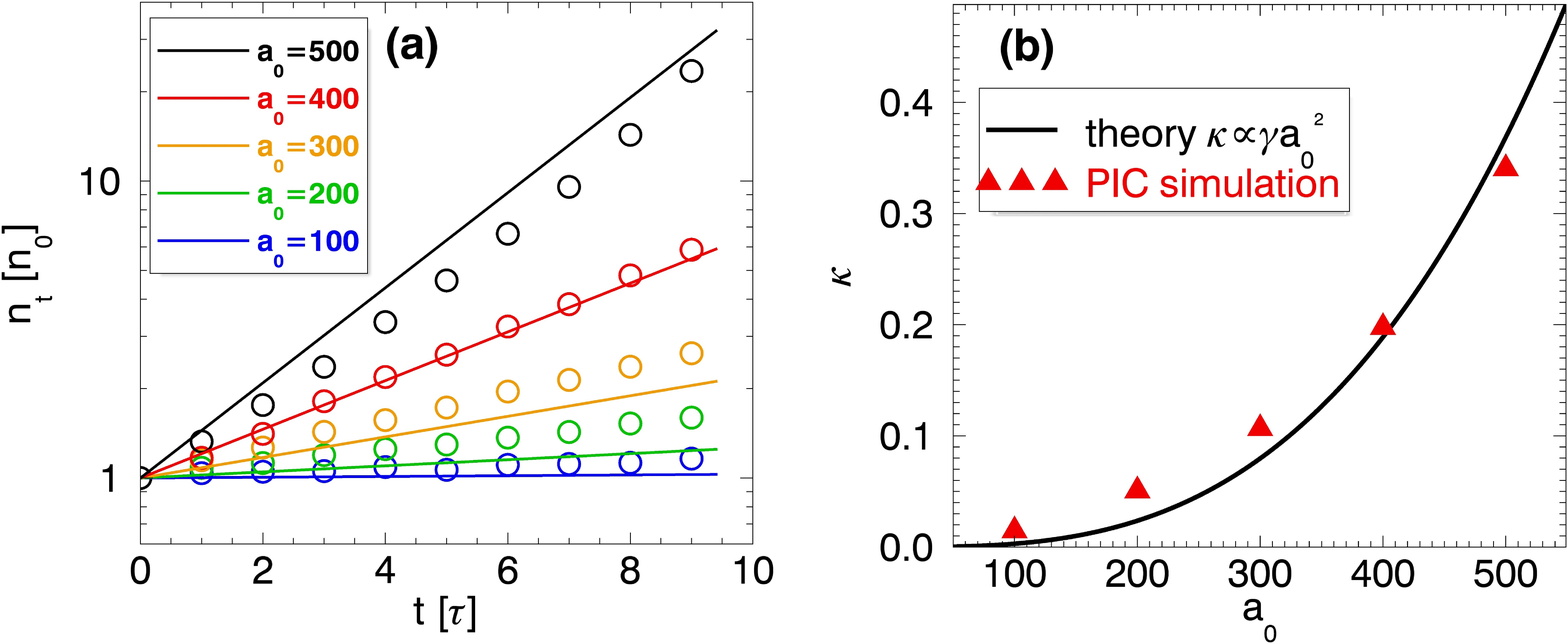}
\caption{(a)The normalized density variation with time at spiral node. Solid line presents the theoretical prediction $n_t=n_0e^{\frac{16\pi r_e}{3\lambda}\gamma a_0^2t}$, circle indicates results from PIC with classic LL reaction force. Different colour correspond to different pulse amplitude $a_0$. (b) solid black line shows the density growth ratio in theoretical form $\kappa=\frac{d\ln n_t}{dt}=\frac{16\pi r_e}{3\lambda}\gamma a_0^2$ while the red triangle denotes counterpart of PIC simulation results}
\label{fig3}
\end{figure}

In addition we record the spiral node density evolution for different amplitude $a_0$ during $0<t<9\tau$, as shown in Fig. \ref{fig3}(a). The theoretical line $n_t=n_0e^{\frac{16\pi r_e}{3\lambda}\gamma a_0^2t}$ is deduced utilizing inverse of contraction ratio $e^{-2|\lambda_+\lambda_-|t}$, where $\gamma$ is estimated by electron initial Hamiltonian $\gamma\sim\sqrt{1+2a_0\frac{2\pi}{\lambda}0.005\lambda}$. When the amplitude of SW field is small enough for $a_0=100,200$, Taylor expansion of exponential growth ratio $e^{\frac{16\pi r_e}{3\lambda}\gamma a_0^2t}\thickapprox1+\frac{16\pi r_e}{3\lambda}\gamma a_0^2t$ indicates density of attractor point $n_t$ is proportional to time $t$. The density growth rates can be defined as:
\begin{equation}
\begin{aligned}\label{eq9}
 \kappa=\frac{d\ln n_t}{dt}=\frac{16\pi r_e}{3\lambda}\gamma a_0^2
 \end{aligned}
\end{equation}
which is illustrated in Fig \ref{fig3} (b) and the simulation results are in good agreement with Eq.\ref{eq9}. The evolution prediction exhibits its exponential growth property where a little increment of $a_0$ lead to substantial density accumulation in attractor. The PIC simulation results demonstrate the attractor density evolution is exactly determined by the RR force format and this may provide us a potential method to detect the LL force formats experimentally.

\begin{figure}[!]
\centering
\includegraphics[width=\textwidth]{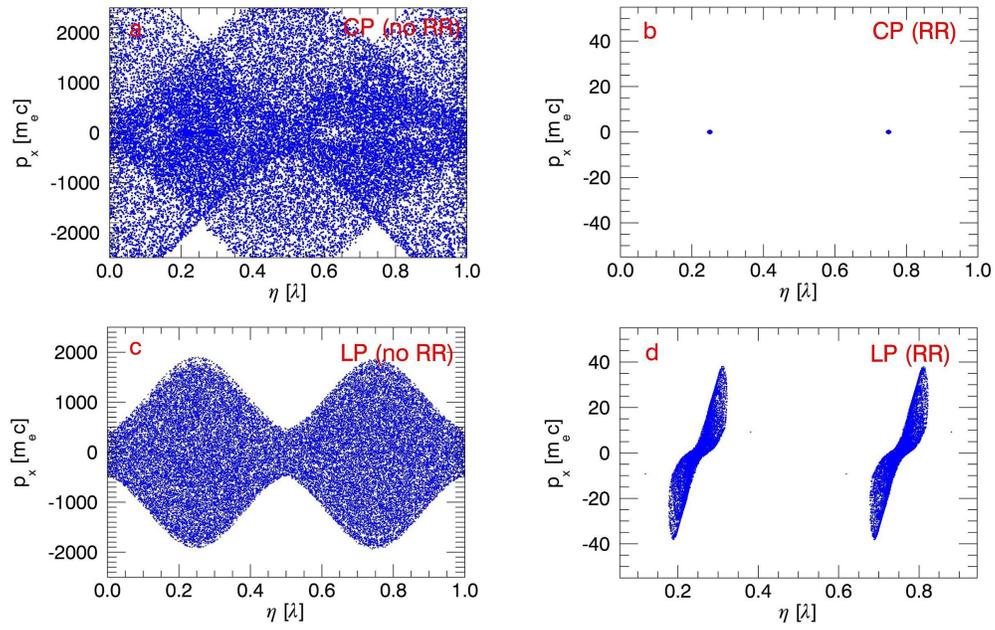}
\caption{Poincar\'{e} plot $p_x$ vs $\eta$ for (a) CP without RR (b) CP with RR and (c) LP without RR (d) LP with RR. Defining $\eta=x$ mod $\lambda$, at times $t=n\tau$, $n=0,1....$. Plots show 100 trajectories for t from 500$\tau$ to 1000$\tau$.}
\label{fig4}
\end{figure}

Above discussion is mainly dedicated to electron's property in vicinity of the attractor. Without loss of generality, it is essential to think over the initial condition $p_\perp|_{t=0}\neq A_\perp|_{t=0}$ where the electron relativistic factor $\gamma$ is no longer conserved. Considering realistic interaction between plasma and colliding laser pulse, we place 100000 electrons whose spatial distribution is random in x-direction and temperature is $50$ MeV to simulate its behaviour under periodic SW field. There both LP ($a_0=300\sqrt{2}$) and CP ($a_0=300$) are included. The Poincar\'{e} plots for different polarizations and with or without RR are shown in Fig. \ref{fig4}. Without RR recoil in Fig. \ref{fig4}(a)(c), both CP and LP demonstrate Hamiltonian chaos and we observe global stochastic behaviour. While taking account of radiation damping, both systems of CP and LP become dissipative and they switch from Hamiltonian chaos\cite{PhysRevLett.88.055004} to attractor behaviour. The comprehensive explanation about LP condition was proposed by G.Lehmann in Ref \cite{lehmann2012phase}. It illustrates that phase space stochastic heating turning into stranger attractor due to RR force in SW field. In RR damping circumstance Fig. \ref{fig4} (b)(d), the phase space contraction phenomenon for CP is much more significant than LP one and most of section points locate in spiral sink node in Fig. \ref{fig4} (b), which indicates the regular stable attractor emerges in $x-p_x$ phase space. Consequently the electrons in Linear polarized SW field can be dispersed away from electric node while regular attractor in CP case induced cooling effect is prone to constrain electrons more tightly, where the spiral sink effect play important role. Meanwhile the $e^-e^+$ pair plasma generation from CP is less efficient than LP due to the dispersed electrons with higher energy and emission probability\cite{M2016Electron}.

\begin{figure}[!]
\centering
\includegraphics[width=\textwidth]{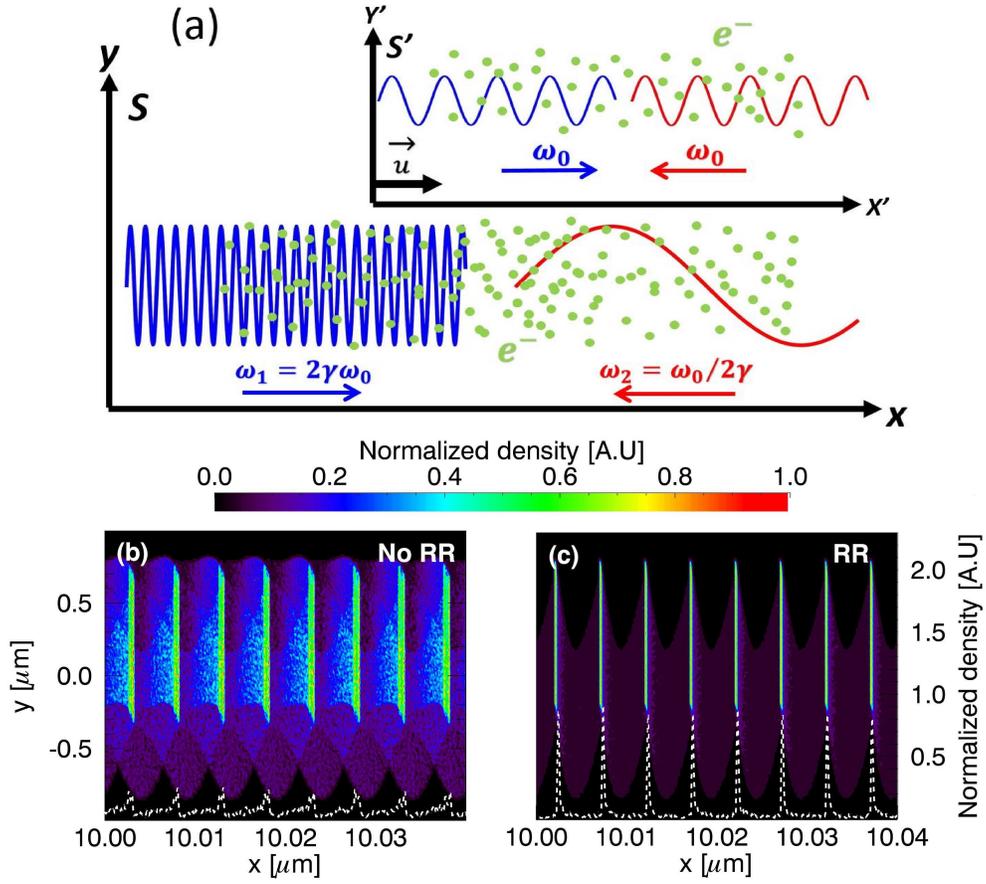}
\caption{(a) Schematic of the Doppler shift of colliding electromagnetic wave from inertial frame to laboratory frame. (b)(c) the snapshot of electron density distribution when considering RR effect or not respectively. The white dash line denotes the value obtained in section of y=0}
\label{fig5}
\end{figure}

\section{Potential application}
\label{Section4}
In addition, we outline some potential applications of spiral attractor caused by RR effect. Suppose that there is a bunch static electron irradiated by CP pulses from both sides in the inertial frame $S'$, as shown in Fig \ref{fig5} (a). The frame $S'$ is moving with velocity $u$ in the positive $x$ direction with respect to the laboratory frame $S$. In 2D PIC simulation, a bunch of electron is initialized with normalized energy $\gamma=100$ moving along the x-direction with velocity $u=\sqrt{1-1/\gamma^2}$ and the CP laser pulses with amplitude $a_0=300$ from left and right boundary are set as infinite plane wave whose circular frequencies are $\omega_1=2\gamma\omega_0$ and $\omega_2=\omega_0/(2\gamma)$ respectively, where $\omega_0$ is correspond to frequency of $1\mu m$ wavelength. The simulation territory is a $20\mu m\times2\mu m$ rectangle divided by $200000\times200$ cells. Utilizing Doppler shift formulas, in $S'$ frame laser frequencies are $\omega_1'=\omega_0$ and $\omega_2'=\omega_0$ as shown in Fig \ref{fig5} (a). A similar simulation as above had been done in frame $S'$ to accomplish periodic electron slices with length $\delta l<1\mu m$ due to spiral sink node attractive effect. In the laboratory frame, these slices length can be calculated as $\delta l/\gamma < 10nm$ through Lorentz transformation, which is in a perfect agreement with result in Fig \ref{fig5} (c), where the electron slice length $\delta l\approx1nm$. When ignoring RR effect, the SW nodes without attractive behaviour cannot confine electrons in small area so that the nanometer-scale electron slice is difficult to realize in Fig \ref{fig5} (b). The parameters of the slices in Fig \ref{fig5} (c) can be determined through adjusting the frequencies $\omega_1$ and $\omega_2$. Supposing $\omega_1>\omega_2$ and $\varepsilon\gg1$, the electron slice normalized directional energy is $\varepsilon=\sqrt{\omega_1/(4\omega_2)}$ and its length is $\delta\ll2\pi c/(\varepsilon\sqrt{\omega_1\omega_2})$. Such a series of ultrashort and nanobunched electron slices can be used to generate coherent transition radiation\cite{PhysRevLett.91.074802} and generate a train of attosecond x-ray pulses through coherent Thomson backscattering\cite{PhysRevLett.104.234801}.

\section{Conclusions and outlook}
Conclusion, we propose a novel analytic method to investigate electron dynamics in ultra-intense counter-propagating laser fields. Utilizing the linear stability theory, we found that spiral attractors emerge in electron phase-space when classical radiation reaction is taken into account. The electrons are spirally approaching the attractors and the electron densities nearby grow exponentially. The growth rates were given theoretically, which are proportional to the laser intensities. Experimentally measuring these growth rates can provide a quantitative way to diagnose the formula of radiation reaction force.
\section*{Acknowledgements}
The work has been supported by the National Basic Research Program of China (Grant No.2013CBA01502), NSFC (Grant Nos.11535001) and National Grand Instrument Project (2012YQ030142). The PIC code Epoch was in part funded by the UK EPSRC grants EP/G054950/1, EP/G056803/1, EP/G055165/1 and EP/M022463/1. Our simulations were carried out in Max Planck Computing and Data Facility and Shanghai Super Computation Center. The author Z.Gong acknowledges useful discussion with S.V. Bulanov and H.X. Chang. Z. Gong and R.H. Hu contributed equally to this work.

\bibliographystyle{model1-num-names}
\bibliography{reference}

\end{document}